\documentclass[oneside,twocolumn,aps,pra,preprintnumbers,bibnotes10pt,superscriptaddress,amsmath,amssymb]{revtex4-2}

\usepackage{graphicx}
\usepackage{subcaption}
\usepackage{float}
\usepackage{color}
\usepackage{epstopdf}
\usepackage{amsmath}
\usepackage{braket}
\usepackage{amsthm}
\usepackage{amssymb}
\usepackage{amsfonts}
\usepackage{txfonts}
\usepackage{mathtools}
\usepackage{bm}
\usepackage{hyperref}
\usepackage{lipsum}
\usepackage{natbib}
\usepackage{comment}
\usepackage{physics}
\usepackage{cancel}
\usepackage{lipsum}

\usepackage{tikz}
\usetikzlibrary{quantikz2,calc,arrows,chains,matrix,positioning,scopes}

\usepackage{letltxmacro}
\LetLtxMacro{\originaleqref}{\eqref}
\renewcommand{\eqref}{Eq.~\originaleqref}

\newcommand{\Time}{\mathcal{T}}

\newcommand{\re}{{\rm Re}}

\def \ket#1{\mathinner{|{#1}\rangle}}
\def \bra#1{\mathinner{\langle{#1}|}}

\newcommand{\Idoperator}{\mathbb{I}}

\newcommand{\Prob}{\mathcal{P}}
\newcommand{\KND}{K}

\newcommand{\MU}{\hat{U}}

\begin{document}

\title{Quantum Non-Demolition Measurements and Leggett-Garg inequality}

\author{Paolo Solinas}
\affiliation{Dipartimento di Fisica, Universit\`a di Genova, via Dodecaneso 33, I-16146, Genova, Italy.}
\affiliation{INFN - Sezione di Genova, via Dodecaneso 33, I-16146, Genova, Italy.}

\author{Stefano Gherardini}
\affiliation{Istituto Nazionale di Ottica del Consiglio Nazionale delle Ricerche (CNR-INO), Largo Enrico Fermi 6, I-50125 Firenze, Italy.}
\affiliation{European Laboratory for Non-linear Spectroscopy, Universit\'a di Firenze, I-50019 Sesto Fiorentino, Italy.}

\begin{abstract}
Quantum non-demolition measurements define a non-invasive protocol to extract information from a quantum system that we aim to monitor.
They exploit an additional quantum system that is sequentially coupled to the system. 
Eventually, by measuring the additional system, we can extract information about temporal correlations developed by the quantum system dynamics with respect to a given observable.
This protocol leads to a quasi-probability distribution for the measured observable outcomes, which can be negative.
We prove that the presence of these negative regions is a necessary and sufficient condition for the violation of macrorealism.
This is a much stronger condition than the violation of the Leggett-Garg inequalities commonly used for the same task.
Indeed, we show that there are situations in which Leggett-Garg inequalities are satisfied even if the macrorealism condition is violated.
As a consequence, the quantum non-demolition protocol is a privileged tool to identify with certainty the quantum behavior of a system. As such, it has a vast number of applications to different fields from the certification of quantumness to the study of the quantum-to-classical transition.
\end{abstract}

\maketitle

\section{Introduction}

How can we distinguish a quantum system from a classical one?
What are the distinctive features of a quantum system? 
These are two of the questions physicists have asked from the dawn of quantum theory. 
To give a clear and definite answer is difficult because of the smooth boundary between the classical and quantum worlds, and due to our inability to directly assess quantum features without altering them.
For example, when a quantum system interacts with an environment, it quickly loses its quantum features and starts to behave classically.
Still, the identification of protocols to spot and quantify a system's quantum features is of paramount importance and it would have a deep impact on both the foundation of quantum mechanics and quantum technologies.

The milestones in this field are the papers by John Bell in $1964$~\cite{Bell1964} and by Leggett and Garg~\cite{Leggett1985} in $1985$.
They both try to answer our initial questions, identifying and giving some criteria to highlight two quantum features.
The famous Bell's inequalities~\cite{Bell1964,Freedman1972,Aspect1982,Pan1998} focus on the non-locality and the quantum correlations related to the entanglement while the Leggett-Garg's inequalities (LGIs), which formally have a similar structure, discuss the violation of the macrorealism (MR)~\cite{Leggett1985}.

The concept of MR was introduced to identify the features that a classical system should have and then test if they are present at a quantum level. If not, one would conclude that a quantum system behaves in a non-classical way.
In the following, for the sake of continuity with the current literature, we first use the original terminology of MR~\cite{Leggett1985}, which is still adopted in several contributions, but then we will give a modern and precise definition~\cite{maroney2014,Schmid2024}.
The Leggett and Garg's assumptions to test the MR are $1)$ Macrorealism per se (MRps), whereby the system is in one of the states available to it at each moment, and $2)$ Noninvasive measurability (NIM) that identifies the conditions under which one can determine the state of a system without disturbing the subsequent dynamics.

Later, an additional condition (implicit in the original article~\cite{Leggett1985}) was identified~\cite{Halliwell2016} as $3)$ Induction: future measurements cannot affect the present state. All these assumptions are surely satisfied by a classical system, while they might be violated for quantum systems.
Therefore, they give a testable way to distinguish between quantum and classical behavior. In this regard, it would be thus desirable to identify a condition that certifies the violation of macrorealism, eventually with certainty.

The LGIs are usually stated in terms of quantum correlators of sequential measurements performed on a single quantum system.
The violation (resp.~validation) of the LGIs is a signature of the quantum (resp.~classical) behavior of the system of interest.
However, the violation of LGIs is only a sufficient condition for the violation of MR~\cite{Halliwell2016}, which is equivalent to stating that the validity of LGIs is a necessary condition such that even the MR is fulfilled.
Practically, this means that there could be situations in which the MR is violated but the LGIs are satisfied.

In this paper, we focus on the MRps condition, comprising the MR, by giving a new general protocol guaranteeing both a necessary and sufficient condition to identify its violation. 
The violation of the MRps implies the violation of the MR's assumption that, in turn, identifies the presence of quantum features. Such a protocol is based on quantum non-demolition (ND) measurements~\cite{Braginsky1980,BraginskyBook, Caves1980, CavesRevModPhys} and it was proposed in Refs.~\cite{solinas2013work,solinas2015fulldistribution,solinas2016probing,solinas2021,solinas2022,GherardiniTutorial}. 
To highlight the practical implications and experimental feasibility of the protocol, we will start by describing its implementation which makes use of a quantum detector to store the desired information. In this way, by measuring the state of the detector, one can construct a quasi-probability distribution (QPD) of the corresponding measurement outcomes. 
As clarified in Refs.~\cite{LostaglioKirkwood2022,GherardiniTutorial} other quasi-probability distributions have been defined in the literature, ranging from Kirkwood-Dirac quasi-probabilities~\cite{yunger2018quasiprobability,ArvidssonShukurJPA2021,DeBievrePRL2021,LostaglioKirkwood2022,BudiyonoPRAquantifying,wagner2023quantum,ArvidssonShukur2024review,hernandezArXiv2024Interfero} to the full counting statistics~\cite{LevitovJMP1996,NazarovEPJB2003,ClerkPRA2011,HoferPRL2016} and Keldysh quasi-probabilities~\cite{Hofer2017quasiprobability,PottsPRL2019}. 
At the same time, alternative protocols to determine the violation of the LGIs with weak measurements have been proposed \cite{Ruskov2006, Jordan2006, Goggin2011}.
Nevertheless, in this paper, our focus goes to quasiprobabilities based on ND measurements, as we are going to prove that the corresponding distribution has negative regions if and only if the MRps condition is violated.

In order to show the effectiveness of our results, we compare the ability of our protocol to reveal a violation of MR with the LGIs, representing so far the proper tool for such a task. For the sake of a fair comparison, we derive, from the full QPD, the multi-time quantum correlators leading to the corresponding LGIs.
With an in-depth analysis, we identify the contributions of the QPD that lead to the violation of the LGIs and we show that only in some particular cases the LGIs spot the violation of the MR, thus confirming that such a circumstance is only a sufficient condition.
To further stress this point and test on a concrete case study, we consider a specific example taken from Ref.~\cite{Halliwell2016}.
We show how, by using the negative regions of the QPD, we are always able to identify any violation of MRps and thus of MR, while LGIs statistically fail in half of the cases.

To the best of our knowledge, the protocol based on ND presented here is the first allowing for the certain identification of purely quantum behaviors associated with the violation of MR.
This feature joined with the fact that it finds application beyond binary observables, for which usually the LGIs are discussed, has implications for the foundation of quantum mechanics and quantum technologies. For example, it can be used (i) to characterize the quantum-to-classical transition~\cite{Arndt1999, Gerlich2011, Aspelmeyer2022, FuchsScienceAdvances2024}, (ii) to monitor the stability and robustness against noise of quantum devices and computers subject to decoherence process~\cite{solinas2021, Kim2023, Bravyi2024}, or (iii) to certify quantum random number generator~\cite{Nath2024}.
Such a reliable and quantitative tool would not only enhance our comprehension of quantum phenomena but also inspire innovative methods for preserving the quantum properties essential to technological advancements.

\section{Macrorealism conditions}
\label{sec:MR_conditions}

The original definition of MRps given by Leggett and Garg  \cite{Leggett1985} was ambiguous and prone to different interpretations.
During the years, this led to confusion about its meanings and implications. Only recently some authors have clarified that it was intended to understand if a quantum system can be described in terms of classical random variables~\cite{maroney2014,Schmid2024}.
More precisely, once the observable to be measured is identified, the MRps assumption is satisfied if {\it quantum superpositions are not allowed, at any time, within the Hilbert space spanned by the eigenbasis of the observable or by a statistical mixture of them}~\cite{Schmid2024}.
In other words, the MRps assumption is satisfied if there cannot be {\it coherent superpositions} and the state of the system is always diagonal in the basis that diagonalizes the observable to be measured~\cite{Schmid2024,Knee2016}.

The NIM assumption, entering MR, is equally difficult to formalize. For this reason, it is usually flanked by the no-signaling in time (NSIT) assumption~\cite{Kofler2008, Halliwell2016}, which is regarded as a statistical (and slightly less strong) version of NIM.
As extensively discussed by Halliwell~\cite{Halliwell2016}, this allows us to relate the NIM with the possibility of attaining the marginal probability of single events from the joint probability of the sequence of them.

That is, as an example, if $p(s_1,s_2,s_3)$ is a generic joint probability of recording $s_1$, $s_2$ and $s_3$, the NSIT condition over, say, the outcome $s_3$, reads: $\sum_{s_1, s_2}  p(s_1, s_2, s_3) = P(s_3)$, where $P(s_3)$ is the probability to measure $s_3$ in correspondence of the state of the system at the time the measurement is performed, ensuring that no previous measurements in the sequence have affected it. For the sake of clarity, throughout the manuscript, the symbol $P$ will always denote probabilities at single times, given by Born's rule, and probabilities at multi-times respecting the theory of classical probability.

Below, we show that the ND quasi-probability always satisfies the NSIT assumption so that the violation of MR can be reduced only to the violation of the MRps' assumption, for which we give both a necessary and sufficient condition.

\section{Three measurement quasi-probability distribution}
\label{sec:three_measurement_QPD}

We consider the following scheme for sequential non-demolition measurements. Suppose we have a quantum system $\mathcal{S}$ evolving under a unitary transformation in a time interval $0 \leq t \leq \Time$. At times $t_0\leq t_1 \leq t_2=\Time $, we measure a generic observable $\hat{A}$ (Hermitian operator).

In general, any projective measurement perturbs a quantum system inducing the collapse of its wave function, thus destroying the quantum coherence between eigenstates of the measured observable. To avoid this drawback in a scheme with sequential measurements, according to the ND approach, we use an auxiliary quantum detector to store the desired information in the phase of the detector's state, which is eventually measured. This scheme allows us to preserve the quantum coherence in the initial density operator and obtain the average value of observables at multi-times, not perturbed by the interaction with the detector~\cite{solinas2015fulldistribution, solinas2016probing,solinasPRA2017, solinas2021,solinas2022}.

Sequential non-demolition measurements rely on a sequence of fast (with respect to the timescale of the system evolution) system-detector couplings. 
The unitary transformations describing the coupling processes are $\hat{u}_k  = \exp\{ i (\lambda/2) \hat{A} \otimes \hat{p} \}$, which occurs at any time $t_k$ with $k=0, 1, 2$ (see Refs.~\cite{solinas2013work,solinas2015fulldistribution,solinas2016probing} for more details). Here, $\lambda$ is an effective coupling, and $\hat{p}$ is an operator acting on the detector's degrees of freedom.

\begin{figure}
    \begin{center}
    \includegraphics[scale=.5]{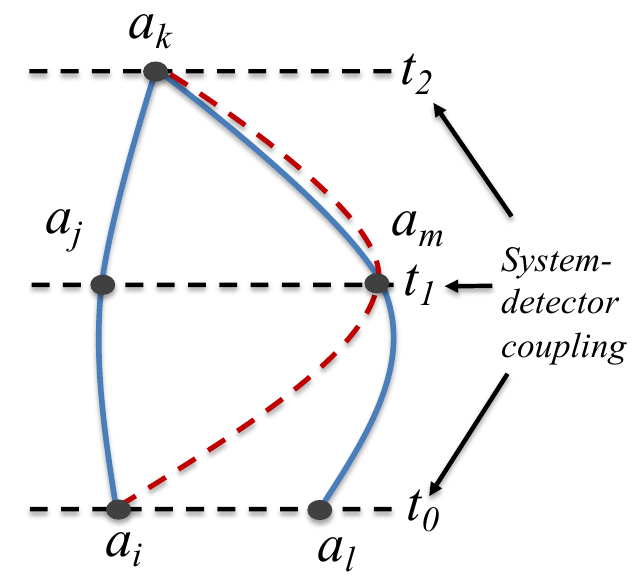}
    \end{center}
    \caption{A pictorial representation of the evolution in terms of sequential measurements at three times, and Feynman paths built over the measurement outcomes.
    The measurements occur at time $t_0$, $t_1$ and $t_2$. The possible outcomes of the measurements are $a_i$ and $a_l$ at time $t_0$, $a_j$ and $a_m$ at time $t_1$, and $a_k$ at time $t_2$. The sequence of outcomes identifies the paths that the monitored system could follow during its evolution.    
    The red dashed curve represents a macrorealistic (or classical) path where, at any time, we have a single outcome (e.g., $a_i \rightarrow a_m \rightarrow a_k$) so that the observable has a determinate outcome.
    The blue curves present quantum paths in which the system is in a superposition of states with different outcomes.
    In this case, the observable $\hat{A}$ does not have a unique value and the paths violated the macrorealism requirement.} 
    \label{fig:superpositions}
\end{figure} 

We denote the unitary transformation from time $t_{j-1}$ to $t_j$ acting on the system only as $\MU_j \otimes \Idoperator = \MU( t_{j}, t_{j-1} ) \otimes \Idoperator$ (with $j=1,2$) so that the total unitary transformation acting on both the system and detector is $\MU_{tot} = \hat{u}_2 \MU_2 \hat{u}_1 \MU_1 \hat{u}_0$.
In the following, to simplify the notation we will denote $ \MU_i \equiv \MU_i \otimes \Idoperator$. Explicitly, the total evolution operator $\MU_{tot}$ reads as
\begin{equation}
\MU_{tot} = e^{ i \frac{\lambda}{2} \hat{A} \otimes \hat{p} } \MU_2 \, e^{ i \frac{\lambda}{2} \hat{A} \otimes \hat{p} } \MU_1 \, e^{ i \frac{\lambda}{2} \hat{A} \otimes \hat{p} }.
\label{eq:U_tot}
\end{equation}
Notice that, since in general $[\hat{U}_j, \hat{A}] \neq 0$, this scheme can also be applied to describe the sequential measurement of three distinct non-commuting operators $\hat{A}$, $\hat{B}$ and $\hat{C}$. Observe also that this ND protocol can be implemented in any quantum platform that allows for a tunable interaction between the quantum system of interest and an auxiliary system acting as the detector. For example, in Ref.~\cite{solinas2021}, the scheme was implemented using several qubits to describe the system and one qubit for the detector.

We decompose the initial state of the system in the basis $\{ \ket{i} \}$ in which the observable $\hat{A}$ is diagonal, i.e.~$\hat{A} \ket{i} = a_i \ket{i}$ with $a_i$ scalars, so that $\ket{\psi_0} = \sum_i  \psi^0_i \ket{i}$. 
Throughout the paper, we assume that $\ket{\psi_0}$ is not an eigenstate of $\hat{A}$.
On the other hand, the state of the detector is taken to be $\ket{\phi_0} = \left(1/\sqrt{M} \right) \sum_p \ket{p}$, where $\hat{p} \ket{p} = p \ket{p}$. The calculation of the evolution of the system and detector together is a straightforward extension of the one presented in Refs.~\cite{solinas2016probing,DeChiara2018}, and the final total state is 
\begin{equation}
     \ket{\Psi} = \frac{1}{\sqrt{M}}\sum_p \sum_{i, j, k} e^{i \frac{\lambda p}{2}(a_k + a_j + a_i)}     U_{2,kj} U_{1,ji} \psi^0_{i} \ket{k} \ket{p}
\end{equation}
where $U_{2,kj} = \matrixel{k}{\MU_2}{j}$ and $U_{1,ji} = \matrixel{j}{\MU_1}{i}$.

The information on the evolution of the monitored system is extracted by measuring the phase accumulated in the detector between two generic states $\ket{\pm p}$.
This is obtained by taking the off-diagonal element of the detector's density operator and tracing out the degrees of freedom of the system. 
Formally, writing the final total density operator as $\hat{R} = \ketbra{\Psi}{\Psi}$ and the density operator of the detector as $\hat{r}^0 = \ketbra{\phi_0}{\phi_0}$, this is $\mathcal{G}_\lambda = 
\Trace\left[ (\Idoperator\otimes \langle p|)R(\Idoperator\otimes |-p\rangle) \right] / \matrixel{p}{r^0}{-p}$.

Normalizing $p$ to one (in some opportune unit), $\mathcal{G}_\lambda$ explicitly reads as
\begin{equation}
    \mathcal{G}_\lambda = \sum_{i, l, j, m, k} e^{i \lambda \left (a_k + \frac{a_j+a_m+a_i+a_l}{2} \right)}   U_{2,kj} U_{1,ji} \rho^0_{il} U^{*}_{1,lm} U^{*}_{2,mk}
\end{equation}
where $\hat{\rho}^0 = \ketbra{\psi_0}{\psi_0}$ is the density operator of the system at the initial time $t_0=0$, and $\rho^0_{il} = \matrixel{i}{\hat{\rho}^0}{l}$.
The function $\mathcal{G}_\lambda$ is also called quasi-characteristic function~\cite{solinas2015fulldistribution, solinas2016probing,solinasPRA2017,solinas2021,solinas2022}.

As usual, the ND quasi-probability distribution is obtained by calculating the inverse Fourier transform of $\mathcal{G}_\lambda$, i.e.,  $\Prob_{ND}(\Delta) = (2 \pi)^{-1} \int \exp\{ -i \lambda \Delta \} \mathcal{G}_\lambda d \lambda$. If we introduce
\begin{equation}\label{eq:stochastic_Delta}
    \Delta_{k, j, m, i, l} = a_k + \frac{ a_j+a_m+a_i+a_l }{2}
\end{equation}
and we use the projectors $\hat{\Pi}_k = \ketbra{k}{k}$ (with $\hat{\Pi}_k^2 = \hat{\Pi}_k$, $\sum_k \hat{\Pi}_k = \Idoperator$ and $\hat{\Pi}_k \hat{\Pi}_j = \delta_{kj}\hat{\Pi}_k$), $\Prob_{ND}(\Delta)$ is
\begin{equation}
    \Prob_{ND}(\Delta) = \sum_{i, l, j, m, k} P_{ND}(k, j, m, i, l) \delta \left[ \Delta - \Delta_{k, j, m, i, l} \right]
	\label{eq:ND_distribution}
\end{equation}
where $P_{ND}(k, j, m, i, l) = U_{2,kj} U_{1,ji}  \rho^0_{il} U^{*}_{1,lm} U^{*}_{2,mk} = \Trace \left[ \hat{\Pi}_k \MU_2 \hat{\Pi}_j \MU_1 \hat{\Pi}_i \hat{\rho}^0 \hat{\Pi}_l \MU^\dagger_1 \hat{\Pi}_m \MU^\dagger_2 \right]$, and $\delta[\cdot]$ denotes the Dirac's delta. Thus, $P_{ND}(k, j, m, i, l)$ is the probability amplitude to arrive in the state $\ket{k}$ following the {\it superposition of paths} $i \rightarrow j \rightarrow k$ and $l \rightarrow m \rightarrow k$ (see Fig.~\ref{fig:superpositions}).

The ND quasi-probability distribution $\Prob_{ND}(\Delta)$ comprises both a classical and a quantum contribution, $\Prob_{\rm cl}(\Delta)$ and $\Prob_{\rm q}(\Delta)$, respectively. The former contribution is the one in which a well-defined value of $\hat{A}$ is recorded {\it at any measurement time} by measuring the system. That is, for $m=j$ and $l=i$, so that
\begin{equation}
    \Prob_{\rm cl}(\Delta) = \sum_{i, j, k} P_{ND}(k, j, j, i, i) \delta \left[ \Delta - \Delta_{k, j, j, i, i}   \right].
\end{equation} 
These can be interpreted as probabilities obeying classical probability theory since $P_{ND}(k, j, j, i, i) = P^{(0)}_{i} P^{(1)}_{i \rightarrow j} P^{(2)}_{j \rightarrow k}$ where $P^{(0)}_i =  \rho^0_{ii}$, $P^{(1)}_{i \rightarrow j} = |U_{1,ji}|^2$ and $P^{(2)}_{j \rightarrow k} = |U_{2,kj}|^2$ are the probabilities of finding the system initially in the state $\ket{i}$, of having a transition from state $i$ to $j$ and from $j$ to $k$, respectively.

The quantum contributions in Eq.~(\ref{eq:ND_distribution}), thus with $m\neq j$ or $l\neq i$, are associated with {\it quantum trajectories} where the system is in a coherent superposition of paths defined by the measurement outcomes; see Fig.~\ref{fig:superpositions}.
More specifically, at time $t_0$, the measurement outcomes might be $a_i$ or $a_l$ while, at time $t_1$, $a_j$ or $a_m$. Formally, $\Prob_{\rm q}(\Delta)$ is
\begin{equation}
     \Prob_{\rm q}(\Delta) = \sideset{}{'}\sum_{i, l, j, m, k} P_{ND}(k, j, m, i, l) \delta \left[ \Delta - \Delta_{k, j, m, i, l}  \right] 
     \label{eq:P_q}
\end{equation}
where the sum $\sideset{}{'}\sum$ does not include the terms for which $m = j$ and $l = i$ entering the classical distribution $\Prob_{\rm cl}(\Delta)$; this means that in $\Prob_{\rm q}(\Delta)$ are allowed contributions with ($l = i$, $m \neq j$), ($m = j$, $l \neq i$) and ($m \neq j$, $l \neq i$).

We stress that the violation of MRps in the $\Prob_{\rm q}(\Delta)$ shall be manifest also in the presence of classically forbidden values of $\Delta_{k, j, m, i, l}$~\cite{solinas2015fulldistribution, solinas2016probing,solinasPRA2017,solinas2021,solinas2022}, as given in \eqref{eq:stochastic_Delta}.
For example, suppose that $a_i = \pm 1$. 
The possible classical (i.e., $m = j$ and $l = i$) values of $\Delta$ are $\{ -3, -1, 1, 3 \}$. On the contrary, the quasi-distribution $\Prob_{\rm q}(\Delta)$ also accounts for terms associated with $\Delta = \pm 2$ (for example, with $a_k=a_j=a_m=a_i=1$ and $a_l=-1$) or $\Delta = 0$ (for example, with $a_k=a_j=1$ and $a_m=a_i=a_l=-1$).

The QPD $\Prob_{ND}(\Delta)$ in \eqref{eq:ND_distribution} satisfies the noninvasive measurability condition, in terms of the no-signaling in time assumption~\cite{Leggett1985,Fritz_2010,Kofler2013, Emary_2014,Halliwell2016}. This means that marginalizing $\Prob_{ND}(\Delta)$ over all the measurement outcomes but the ones at a single time returns the corresponding probability to measure the observable $\hat{A}$ given by the Born's rule. 
For example, to obtain the probability of recording a final eigenvalue $a_k$, we must sum over all the intermediate outcomes $a_i$, $a_l$, $a_j$, $a_m$. It can be shown (see Methods) that  
$P(a_k) = \sum_{i, l, j, m} P_{ND}(k, j, m, i, l) = \Trace \left[ \hat{\Pi}_k \hat{U} \hat{\rho}^0 \hat{U}^\dagger \right]$.
Indeed, $P(a_k)$ is the probability to measure $\hat{A}$ at time $\Time$, when the system has evolved under the unitary transformation $\hat{U}$~\cite{Halliwell2016}.
Similarly, the probability to record the eigenvalues $a_i$ and $a_l$ at time $t=0$, is obtained from \eqref{eq:ND_distribution} by summing over the indices $l, j, m, k$ and $i, j, m, k$ respectively. Thus, we have $P(a_i) = \sum_{l, j, m, k}P_{ND}(k, j, m, i, l) = \Trace \left[ \hat{\Pi}_i \hat{\rho}^0 \right]$ as expected.
With similar calculations, it can be shown that also the intermediate measurements satisfy the NIM condition, such that $P(a_j) = \sum_{i, l, m, k} P_{ND}(k, j, m, i, l) = \Trace \left[ \hat{\Pi}_j \hat{U}_1 \hat{\rho}^0 \hat{U}_1^\dagger \right]$ and $P(a_m) = \sum_{i, l, j, k} P_{ND}(k, j, m, i, l) = \Trace \left[ \hat{\Pi}_m \hat{U}_1 \hat{\rho}^0 \hat{U}_1^\dagger \right]$. 
It is also worth observing that summing $P_{ND}(k, j, m, i, l)$ over three indices returns the Kirkwood-Dirac quasi-probabilities at two-times~\cite{LostaglioKirkwood2022,GherardiniTutorial,ArvidssonShukur2024review}, apart the 3-tuples $(j,m,k)$ and $(i,l,k)$ that give rise to a single-time probability. By construction, Kirkwood-Dirac quasiprobabilities respect the assumption of no-signaling in time.

\section{Negative regions are necessary and sufficient for violating macrorealism}
\label{sec:necessary_sufficient_condition}

We now prove that $\Prob_{ND}(\Delta)$ has negative contributions if and only if the macrorealism, i.e., MRps, condition is violated. The proof is composed of three steps.
First, we show that $\Prob_{ND}(\Delta)$ is real, i.e.~given by a distribution of real numbers. Then, we prove that it is normalized to $1$ and that this normalization comes uniquely from the classical contribution $\Prob_{\rm cl}(\Delta)$. This implies that the terms entering the quantum contribution $\Prob_{\rm q}(\Delta)$ cancel out each other and that some of them must be negative. The last part of the proof involves the connection between the macrorelism and the negativity of $\Prob_{ND}(\Delta)$.

To prove that $\Prob_{ND}(\Delta)$ is real, we group the terms in the distribution that are multiplied by the same $\delta$-function $\delta[\Delta - \Delta_{k, j, m, i, l}]$. There are indeed identical values of $\Delta_{k, j, m, i, l}$ because of the symmetries between the exchange of the indices $(i,l)$ and $(j,m)$. Formally, using the properties of the matrix elements, e.g., $U_{2,km} = \matrixel{k}{\MU_2}{ m} =  \left( \matrixel{m}{\MU_2}{ k} \right)^*$, we have $P_{ND}(k, m, j, i, l) = \left( P_{ND}(k, j, m, i, l) \right)^*$ so that \eqref{eq:P_q} can be rewritten as 
\begin{equation}\label{eq:P_q_2}
	\Prob_{\rm q}(\Delta) = 2 \sideset{}{''}\sum_{i, l, j, m, k} \re \left[P_{ND}(k, j, m, i, l) \right]  \delta \left[ \Delta - \Delta_{k, j, m, i, l} \right]	
\end{equation}
where $\sideset{}{''}\sum$ sum over half the terms than $\sideset{}{'}\sum$ as it includes terms for which $l > i$ or $m > j$.
Since the distribution $\Prob_{\rm cl}(\Delta)$ contains (positive) classical joint probabilities by construction and $\Prob_{\rm q}(\Delta)$ comprises real numbers, we conclude that $\Prob_{ND}(\Delta)$ is real.

The proof that $\Prob_{\rm cl}(\Delta)$ is normalized comes from noting that its contributions are positive joint probabilities obeying the classical probability theory. In fact, we have that $\int d \Delta \Prob_{\rm cl}(\Delta) = \sum_{i, j, k} P^{(0)}_{i}  P^{(1)}_{i \rightarrow j} P^{(2)}_{j \rightarrow k} = \sum_{i, j, k} \Trace \left[ \hat{\Pi}_k \MU_2 \hat{\Pi}_j \MU_1 \hat{\Pi}_i \hat{\rho}^0 \hat{\Pi}_i \MU^\dagger_1 \hat{\Pi}_j \MU^\dagger_2 \right] = 1$.

Regarding the integration of the quantum contribution $\Prob_{\rm q}(\Delta)$ of the ND quasi-distribution $\Prob_{ND}(\Delta)$, it is worth distinguishing two distinct sums of terms: one with $m \neq j$, and the other with $(m = j, l \neq i)$. Then, as above, we integrate both sums of terms in $\Prob_{\rm q}(\Delta)$ over $\Delta$.
The integration over $\Delta$ of the sum with $m \neq j$ leads to
\begin{eqnarray}\label{eq:first_part_Pq_Delta}
    2 \sum_{k, i, l, j, m>j} \re \, \Trace \left[ \hat{\Pi}_k \MU_2 \hat{\Pi}_j \MU_1 \hat{\Pi}_i \hat{\rho}^0 \hat{\Pi}_l \MU^\dagger_1 \hat{\Pi}_m \MU^\dagger_2 \right] =0,
\end{eqnarray}
where we have used the projectors' properties $\hat{\Pi}_k\hat{\Pi}_j = \delta_{k,j}\hat{\Pi}_k = 0$ and $\sum_k \hat{\Pi}_k = \Idoperator$.
An analogous result is obtained from integrating over $\Delta$ the sum of terms in $\Prob_{\rm q}(\Delta)$ with $(m = j, l \neq i)$. In fact, one can get that
\begin{equation}\label{eq:second_part_Pq_Delta}
    2  \sum_{k, i, j, l>i} \re \, \Trace \left[ \hat{\Pi}_k \MU_2 \hat{\Pi}_j \MU_1 \hat{\Pi}_i  \hat{\rho}^0 \hat{\Pi}_l \MU^\dagger_1 \hat{\Pi}_j \MU^\dagger_2 \right] = 0 \,.
\end{equation}

Therefore, the normalization of $\Prob_{ND}(\Delta)$ uniquely comes from the classical distribution $\Prob_{\rm cl}(\Delta)$. 
Given that the terms in $\Prob_{\rm q}(\Delta)$ are real numbers and cancel out upon integration, it follows that at least some of them must be negative. Hence, the negativity of $\Prob_{ND}(\Delta)$ derives from contributions in $\Prob_{\rm q}(\Delta)$ that are built with the coherent superposition of $\hat{A}$'s eigenstates at times $t_0, t_1, t_2$. We can thus state that {\it necessary condition for the negativity of $\Prob_{ND}(\Delta)$ is the presence of contributions in $\Prob_{\rm q}(\Delta)$ coming from the coherent superposition of eigenstates associated to an observable that is measured at multiple times}.

Notably, using ND quasi-probabilities, the statement above is also a {\it sufficient condition}, since the negativity of $\Prob_{ND}(\Delta)$ can only derive from $\Prob_{\rm q}(\Delta)$.
Since the two summations in (\ref{eq:first_part_Pq_Delta})-(\ref{eq:second_part_Pq_Delta}) must always be zero, whenever there is a coherent superposition of the measurement observable's eigenstates, i.e., $\Prob_{\rm q}(\Delta)\neq 0$, $\Prob_{ND}(\Delta)$ must contain negative (real) terms. That is, {\it a sufficient condition for the negativity of $\Prob_{ND}(\Delta)$ is the presence of the coherent superposition of eigenstates associated with the observable measured at multiple times}.

We conclude by noting that if a system is in a coherent superposition of the eigenstates of an observable measured at distinct times, then the MRps assumption is violated. Accordingly, {\it the MRps assumption is violated if and only if the ND quasi-probability distribution $\Prob_{ND}(\Delta)$ exhibits negativity}.

\section{Leggett-Garg inequalities}
\label{eq:LGI_violation}

Having established that negative regions in the ND quasi-probability distribution are necessary and sufficient indicators of the violation of MRps, we now compare our results with the usual tool employed so far to identify such a quantum feature: the Leggett-Garg inequalities.

The LGIs are usually formulated for the so-called binary observables~\cite{Leggett1985,Fritz_2010,Emary_2014,Halliwell2016}, that is, for observables that can have only two outcomes.
Hence, while in the previous discussions, the observable $\hat{A}$ had a generic discrete spectrum, here we assume that its eigenvalues can be only $a_i = \pm 1$ for any time $t_i$. LGIs can be built using quantum correlators at two times, say $t_i$ and $t_j$, which are defined as~\cite{Fritz_2010, Emary_2014,Halliwell2016}
$C_{ij} =  \Trace \left[ \left (\hat{A}(t_i) \hat{A}(t_j) + \hat{A}(t_j) \hat{A}(t_i) \right) \hat{\rho}^0\right]/2$, where $\hat{A}(t_i) = \MU^\dagger(t_i, 0)\hat{A}\MU(t_i, 0)$.
The expression of $C_{ij}$ can be rewritten as~\cite{Fritz_2010,Emary_2014,Halliwell2016}
\begin{equation}
	C_{ij} = \sum_{i,j} a_i a_j P(a_j,a_i)  = \sum_{i, j, k} a_i a_j P(a_k, a_j , a_i)
	\label{eq:C_ij}
\end{equation}
where, by construction, $P(a_j, a_i) = \sum_k P(a_k, a_j, a_i)$ and, as above, $P(a_k, a_j, a_i)=P^{(0)}_{i} P^{(1)}_{i \rightarrow j} P^{(2)}_{j \rightarrow k}$ is the joint probability to record the measurement outcomes $a_k$, $a_j$ and $a_i$ at times $t_i,t_j,t_k$ by means of a sequential scheme.  The sequential measurements are performed at three times $t_0=0$, $t_1$ and $t_2$.

Following Ref.~\cite{Emary_2014}, we can calculate the Leggett-Garg (LG) parameter 
\begin{equation}
    K \equiv C_{01} + C_{12} - C_{02}.
\end{equation}
Notably, for any classical system fulfilling MR, the LG parameter $K$ reads as
\begin{equation}\label{eq:K_classical}
    K = 1 - 4 \left[ P(1,-1,1) + P(-1, 1,-1) \right]
\end{equation}
that results in the LGI $-3 \leq K \leq 1$. That is, if the $-3 \leq K \leq 1$ is not satisfied, then at least a condition of MR is violated.

In this context, it would be desirable to find a decomposition of $K$ made of two terms: one equivalent to the right-hand-side of \eqref{eq:K_classical} (valid under MR), and the other different from zero whenever the LGI $-3 \leq K \leq 1$ is violated. For this purpose, we can still resort to the ND quasiprobabilities given that, for binary observables (here, the eigenvalues are $\pm 1$), the correlators $C_{ij}$ can be equivalently expressed as~\cite{Halliwell2016} 
\begin{equation}
C_{ij} = \sum_{i,j} a_i a_j P_{ND}(j, i),
\end{equation}
where $P_{ND}(j, i) = \sum_{l, m, k} [ P_{ND}(k, j, m, i, l) +  P_{ND}(k, m, j, i, l) + P_{ND}(k, j, m, l, i) +  P_{ND} (k, m, j, l, i) ]$ with $P_{ND}(j, i)$ given by Kirkwood-Dirac quasiprobabilities as commented before. Accordingly,
\begin{eqnarray}\label{eq:C01}
    &C_{01} = \displaystyle{ \frac{1}{4} \sum_{i, j} a_i a_j \sum_{l, m, k} \big[ P_{ND} (k, j, m, i, l) +  P_{ND} (k, m, j, i, l) } & \nonumber \\
    &+ P_{ND} (k, j, m, l, i) +  P_{ND} (k, m, j, l, i) \big].&    
\end{eqnarray}
Writing the correlator at two-times $C_{01}$ as a function of the three-time ND quasiprobabilities brings the advantage of including also the superposition of the observable $\hat{A}$'s eigenstates resulting in multiple different possible outcomes: $a_i$ and $a_j$ at times $t_0=0$, and $a_j$ and $a_m$ at time $t_1$.

The definition of the correlators $C_{12}$ and $C_{02}$ involving the (final) measurement at time $t_2$ are similar to that of $C_{01}$: $C_{12} = \sum_{j, k} a_k a_j P_{ND}(k, j)$ and $C_{02} = \sum_{i, k} a_k a_i P_{ND}(k, i)$, whose extended expression is in the Methods.

Since we are working with binary variables, each of them can take two values; in our case, $a_i$ and $-a_i$. For simplicity, we refer to them with the indices $i$ and $\bar{i}$; for example, $P_{ND} (\bar{k}, j, m, i, \bar{i})$ shall stand for $P_{ND}(-a_k, a_j, a_m, a_i, -a_l)$.
Using the correlators $C_{ij}$ as defined above, the LG parameter $K$ for the ND procedure reads as
\begin{eqnarray}\label{eq:K_ND}
    K &=&  \frac{1}{4} \sum_{i, l, j, m, k} f(k, j, m, i, l) P_{ND}(k, j, m, i, l)
\end{eqnarray}
with $f(k, j, m, i, l) = (a_i +a_l)(a_j +a_m) + a_k (a_j +a_m-a_i -a_l)$; see Methods for details. 
It is convenient to separate in $K$ the classical contributions when $m=j$ and $l=i$, i.e.,
\begin{equation}\label{eq:K_cl}
         \KND_{cl} = \frac{1}{2} \sum_{k, j, i} \left[ a_i a_j  + a_k (a_j -a_i ) \right] P_{ND}(k, j, j, i, i),
\end{equation}
from the quantum ones when $m=j$ and $l=\bar{i} \neq i$ 
\begin{equation}
 	\KND_{q,1} = \frac{1}{2} \sum_{i, j, k} a_k a_j  P_{ND}(k, j, j, i, \bar{i})
	\label{eq:K1_quant}
\end{equation}
and when $m=\bar{j} \neq j$ and $l= i$ 
\begin{equation}\label{eq:K2_quant}
    \KND_{q,2} = - \frac{1}{2} \sum_{i, j, k}  a_k a_i P_{ND}(k, j, \bar{j}, i, i).	
\end{equation}

First, we now show that from the classical contributions in \eqref{eq:K_ND} we recover the classical results in the right-hand-side of \eqref{eq:K_classical}.
To do this, we recall that $P_{ND}(k, j, j, i, i) = P^{(0)}_i  P^{(1)}_{i \rightarrow j} P^{(2)}_{j \rightarrow k}$ with $\sum_{i, j, k}P^{(0)}_i  P^{(1)}_{i \rightarrow j} P^{(2)}_{j \rightarrow k}=1$, namely the total probability over a complete set of classical paths must sum to $1$. The latter can be written explicitly as 
$\sum_{i, k} P_{ND}(k, i, i, i, i) + \sum_{k} P_{ND}(k, k, k, \bar{k}, \bar{k}) +\sum_{k}  P_{ND}(k,  \bar{k}, \bar{k}, k, k) = 1$ where we have split the contributions for $j=i$ (and any $k$), $j=k$ {\it and} $i = \bar{k} \neq k$, and $j = \bar{k} \neq k$ {\it and} $i = k$. Using this separation criterion to \eqref{eq:K_cl}, the classical contribution $\KND_{cl}$ can be written as
\begin{equation}
    \KND_{cl} = 1 - 4 \sum_{k} P_{ND}(k, \bar{k}, \bar{k}, k, k).
\end{equation}
Expanding the sum for $k = \pm 1$ leads to the result in \eqref{eq:K_classical} so that $\KND_{cl}$ satisfies the LGI, i.e., $-3 \leq \KND_{cl} \leq 1$.

\begin{figure*}
    \begin{center}
    \includegraphics[scale=.6]{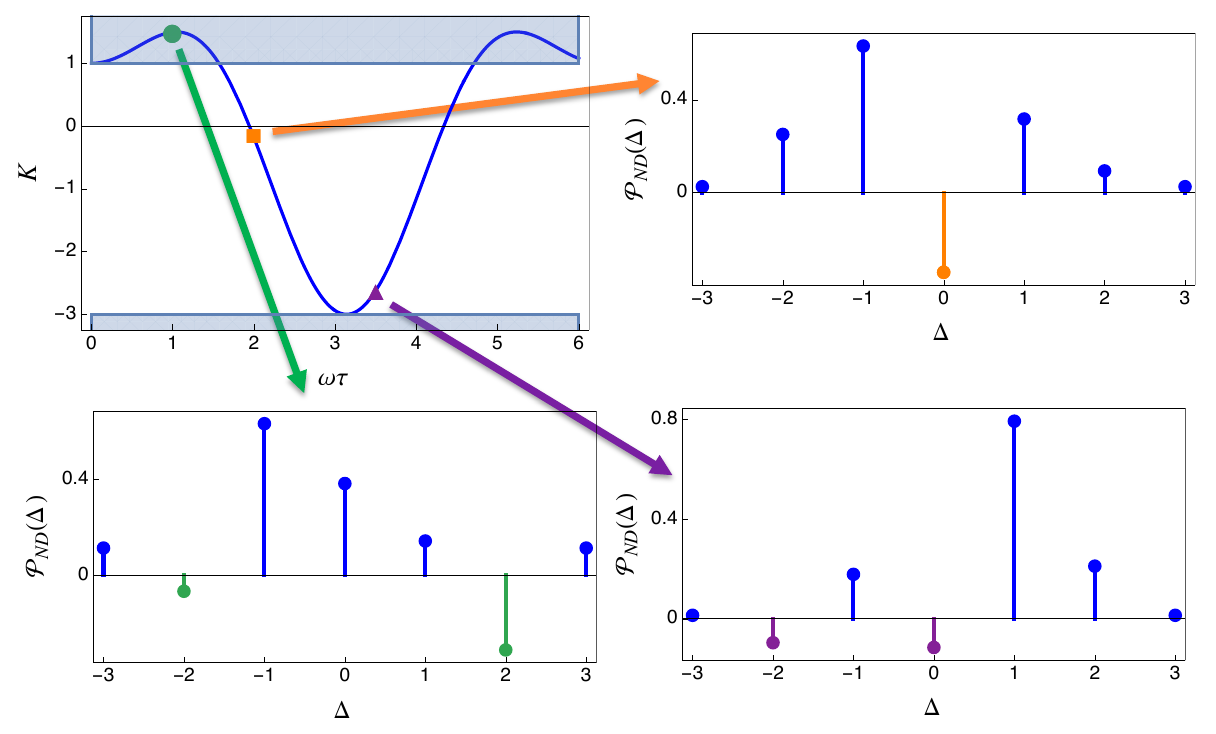}
    \end{center}
    \caption{The top-left panel shows the LG parameter $K$ in \eqref{eq:K} as a function of $\omega \tau$. The shaded regions are the ones for which the LGI is violated. The other panels show the quasi-probability distribution $\Prob_{ND}(\Delta)$ evaluated at three different values of the $\omega\tau$.
    The $\Prob_{ND}(\Delta)$ always shows negative regions, while only for the green circle point the LGI is violated.} 
    \label{fig:LG_VS_PD}
\end{figure*} 

As shown in the Methods, the remaining terms in \eqref{eq:K_ND} can be simplified by noting that $\KND_{q,1} = 0$. Therefore, when present, the violation of the LGI is due to the second quantum contribution $\KND_{q,2}$ in \eqref{eq:K2_quant}.
$\KND_{q,2}$ can be further simplified by observing that $\sum_{k} P_{ND}(k, j,\bar{j}, i, i) = 0$ so that $P_{ND}(\bar{k}, j, \bar{j}, i, i) = - P_{ND}(k, j, \bar{j}, i, i)$. In this way, setting $a_k=1$ without loss of generality, \eqref{eq:K2_quant} equals to
\begin{equation}\label{eq:final_K_q2}
    \KND_{q,2} =-4 \sum_{k} \re \left[ P_{ND}(k, j, \bar{j}, k, k) \right].
\end{equation}
The derivation of \eqref{eq:final_K_q2} is in the Methods.

Some remarks are due to this point.
Using the ND quasi-probability distribution highlights the connection between the violation of macrorealism and LGIs.
In particular, the contributions with $m = \bar{j} \neq j$ and $l=i$, denoting the superposition between eigenstates of $\hat{A}$, are the ones responsible for the violation of the LGI. However, there are situations where the MRps condition is violated but the LGI is satisfied.
This occurs for any initial condition for which $\KND_{q,2}$ vanishes. In such a case, the total quantum contribution $\KND_{q}$ vanishes as well, $\KND_{q} = 0$, with the result that $\KND = \KND_{cl}$ and the LGI is satisfied. Nevertheless, we can have that $l \neq i$, which means that the system is initially in a superposition of eigenstates of $\hat{A}$ and this fact naturally violates the MRps.
On the contrary, if we use the ND quasi-probability $\Prob_{ND}(\Delta)$, the terms $P_{ND}(k, j, j, i, \bar{i})$ in \eqref{eq:K1_quant} always contribute to the negative regions of the quasi-probability distribution. This is a confirmation that the LGIs give only a sufficient condition for the violation of the MRps and thus of the MR's assumption, while the ND quasi-probability $\Prob_{ND}(\Delta)$ gives both a sufficient and necessary condition.

\section{Example}

Following Ref.~\cite{Halliwell2016}, we discuss a specific example that clarifies the results obtained.
Let us consider a two-level quantum system that is initially in the state $\ket{\psi_0} = (\ket{\uparrow} + i \ket{\downarrow})/\sqrt{2}$ where $\ket{\uparrow}$ and $\ket{\downarrow}$ are the eigenstates of the Pauli operator $\hat{\sigma}_z$.
The dynamics is generated by the Hamiltonian $\hat{H} = \omega\hat{\sigma}_x/2$ and we measure the operator $\hat{A}=\hat{\sigma}_z$ at times $t_0=0$, $t_1=\tau$ and $t_2=2\tau$.

The correlators $C_{ij}$ [see \eqref{eq:C_ij}] and the corresponding LG parameter $K$ can be calculated analytically:
\begin{equation}
	K = C_{01} + C_{12} - C_{02} = 2  \cos(\omega\tau) - \cos(2 \omega \tau).
	\label{eq:K}
\end{equation}
As discussed above, for a classical system $-3 \leq K \leq 1$. Hence, any value of $K$ outside this range implies the violation of such an inequality. If we consider the ND protocol, the whole dynamics of the system and the detector is given by \eqref{eq:U_tot}, which leads to the ND quasi-probability distribution in \eqref{eq:ND_distribution}.

The numerical results for the example are shown in Fig.~\ref{fig:LG_VS_PD}.
The top-left panel shows the value of $K$ as a function of $\omega \tau$ with the shaded regions representing the situations in which the LGI $-3 \leq K \leq 1$ is violated.
The colored shapes (dot, square, triangle) give the value of $K$ for three specific values of $\omega\tau$. The other panels of the figure show the ND quasi-probability distribution [as obtained from \eqref{eq:ND_distribution}] that are associated with each colored shape. As we can see, the LGI is violated only for some parameter choices, e.g., the green dot in the top-left panel.
However, for intermediate values of $\omega \tau$ (the orange square and purple triangle), the LGI is satisfied but the dynamics exhibit quantum features given by the presence of quantum coherence along the eigenbases of $\hat{A}$. As proved above, this circumstance is identified by determining negative regions of the corresponding distribution $\Prob_{ND}(\Delta)$.
As expected, the ND quasi-probability distribution is always able to identify the quantum features of the system even when the LGI fails.

From \eqref{eq:K}, the violation of the LGI $-3 \leq K \leq 1$ occurs for $0 \leq \omega \tau \leq \pi/2$ and $3 \pi/2 \leq \omega \tau \leq 2 \pi$. This means that the LGI correctly identifies the violation of MRps in only half of the cases, while the ND protocol always succeeds.

\section{Conclusions}

We have presented a protocol that allows us to identify the presence of genuinely quantum behaviors unambiguously, in terms of the violation of macrorealism per se (MRps).
This protocol is based on performing sequential quantum non-demolition measurements of a given observable $\hat{A}$, and this can be attained using a quantum detector. Using the corresponding measurement outcomes recorded at multiple times, we can construct a quasi-probability distribution. We have proven that the presence of negative regions in such a quasi-probability distribution is a necessary and sufficient condition for the violation of the MRps (and macrorealism) that outlines the presence of quantum behaviors.

The non-demolition protocol has some additional features that make it interesting for more practical applications.
First, it can provide results overcoming some limitations imposed by the LGIs, which are usually employed for binary observables, as the protocol with ND measurements can be implemented for observables with arbitrary discrete spectra. Secondly, being based on an experimental protocol, the ND protocol gives an operational procedure that, in principle, could be implementable on any quantum platform~\cite{solinas2021}.

The advantage of the ND protocol over the LGIs ultimately is a consequence of the more information we extract from the system and its dynamics.
While using LGIs we get information only about correlators at two times, with the ND protocol we have access to a full quasi-probability distribution and, therefore, to all their moments~\cite{solinas2015fulldistribution, solinas2016probing, solinasPRA2017}.
While it is natural to think that the ND protocol gains more information about non-classicality and, thus, needs more resources than other methods as LGIs, the comparison in terms of resources is not straightforward.
Regarding the ND quasi-probability distribution, it is obtained by a Fourier transform of the measured data. Therefore, the resources needed to implement the ND protocol depend critically on the requested precision in performing the Fourier transform. To evaluate this a more detailed analysis is needed.

The research fields that can benefit mostly from the implementation of the ND protocol as a tool to identify quantum behavior are the foundation of quantum mechanics and quantum technologies.
A few examples are the generation of certified random numbers~\cite{Nath2024}, the study of quantum gravity in mesoscopic systems, and the cooling of optomechanics systems up to a scale where quantum effects play a role~\cite{FuchsScienceAdvances2024}.
In all these cases, the question of when and how the quantum-to-classical transition occurs is of critical importance, and it might help to improve already-performed experimental results~\cite{Arndt1999,Gerlich2011} or to better identify causes leading to decoherence in open quantum systems, quantum computers or quantum annealers~\cite{solinas2021, Kim2023, Bravyi2024}.
This could be possible by showing that a given observed quantum-to-classical transition corresponds to a reduction of negative regions in an ND quasi-probability distribution that is built over a properly chosen observable $\hat{A}$. The results in Refs.~\cite{solinas2021,solinas2022} seem to suggest that such a circumstance could be in principle attainable. 

\appendix

\section{Non-invasive measurement}

To compute the probability of measuring an outcome $a_k$ (eigenvalue of $\hat{A}$) from the measurement at time $t_2$, we sum over all the intermediate outcomes $a_i$, $a_l$, $a_j$, $a_m$. In this way, using the properties of the projectors ($\hat{\Pi}_k^2 = \hat{\Pi}_k$, $\sum_k \hat{\Pi}_k = \Idoperator$ and $\hat{\Pi}_k \hat{\Pi}_j = \delta_{kj}\hat{\Pi}_k$), we get 
\begin{eqnarray*}
    P(a_k) &=& \sum_{i, l, j, m} P_{ND}(k, j, m, i, l) \nonumber \\
	&=& \sum_{i, l, j, m} \Trace \left[ \hat{\Pi}_k \MU_2 \hat{\Pi}_j \MU_1 \hat{\Pi}_i \hat{\rho}^0 \hat{\Pi}_l \MU^\dagger_1 \hat{\Pi}_m \MU^\dagger_2 \right] \nonumber \\
	&=& \Trace \left[ \hat{\Pi}_k \MU \hat{\rho}^0 \MU^\dagger \right]  = \Trace \left[ \hat{\Pi}_k \hat{\rho}(\Time) \right]  
\end{eqnarray*}
with $\hat{U} = \MU_2 \MU_1$ and $\hat{\rho}(\Time) = \MU \hat{\rho}^0 \MU^\dagger$. 
$P(a_k)$ is the probability to record $a_k$ at time $\Time$ as given by the Born's rule, consistently with Ref.~\cite{Halliwell2016}.

Concerning the measurement at time $t=0$ of the ND procedure, we have to sum over two possible realizations built with the indices $i$ and $l$. Thus, renaming the indices, the probability to measure the eigenvalue $a_i$ at time $t=0$, can be written as   
\begin{eqnarray*}
    P(a_i) &=& \frac{1}{2}\sum_{l, j, m, k} \Big\{ P_{ND}(k, j, m, i, l) + P_{ND}(k, j, m, l, i) \Big\} \nonumber \\
    &=& \frac{1}{2} \sum_{l, j, m, k} \Big\{  \Trace \left[ \hat{\Pi}_k \MU_2 \hat{\Pi}_j \MU_1 \hat{\Pi}_i \hat{\rho}^0 \hat{\Pi}_l \MU^\dagger_1 \hat{\Pi}_m \MU^\dagger_2 \right] + \nonumber \\
    &+&  \Trace \left[ \hat{\Pi}_k \MU_2 \hat{\Pi}_j \MU_1 \hat{\Pi}_l \hat{\rho}^0 \hat{\Pi}_i \MU^\dagger_1 \hat{\Pi}_m \MU^\dagger_2 \right] \Big\}\nonumber \\
    &=& \Trace \left[ \hat{\Pi}_i \hat{\rho}^0 \right]. 
\end{eqnarray*}
In a similar fashion, we obtain the probability $P(a_l) = \frac{1}{2}\sum_{i, j, m, k} \{ P_{ND}(k, j, m, i, l) + P_{ND}(k, j, m, l, i) \}$.

The calculation for the measurement at time $t_1$ of the ND procedure follows a common reasoning than the one used to get $P(a_i)$. In this case, indeed, there are two possible realizations over the indices $j$ and $m$, such that 
\begin{eqnarray*}
    P(a_j) &=& \frac{1}{2}\sum_{i, l, m, k} \Big\{  P_{ND}(k, j, m, i, l) + P_{ND}(k, j, m, l, i) \Big\} \nonumber \\
	&=& \Trace \left[ \hat{\Pi}_j \MU_1 \hat{\rho}^0  \MU^\dagger_1 \right] = \Trace \left[ \hat{\Pi}_j \hat{\rho}^0(t_1) \right]
\end{eqnarray*}
that, as expected, corresponds to the probability of recording the outcome $a_j$ after the system has evolved to the state $\hat{\rho}^0(t_1)$.
Observing that in general the sum of $P_{ND}(k, j, m, i, l)$ over three indices returns the Kirkwood-Dirac quasiprobabilities at two times, we can conclude that the quasi-probability distribution $\Prob_{ND}(\Delta)$ satisfies the non-invasive measurability condition in terms of the assumption of no-signaling in time.

\section{Correlators and LG parameter}

The two-time quantum correlators $C_{12}$ and $C_{02}$, defined in the main text, explicitly read as
\begin{eqnarray*}
	C_{12} &=& \sum_{j, k} a_k a_j \sum_{i, l, m} \left[  P_{ND} (k, j, m, i, l) + P_{ND} (k, m, j, i, l) \right] \\
	C_{02} &=& \sum_{i, k} a_k a_i \sum_{l, j, m} \left[  P_{ND} (k, j, m, i, l) + P_{ND} (k,  j, m, l,i) \right].
		\label{eq:C12_C_02} 
\end{eqnarray*}
In the expressions of $C_{12}$ and $C_{02}$, we can rearrange the indices to factorize a common $P_{ND}(k, j, m, i, l)$. For example, we can rewrite $C_{12}$ by changing the indices $j \leftrightarrow m$ in $P_{ND}(k, m, j, i, l)$, so that 
\begin{equation*}
	C_{12} = \sum_{i, l, j, m, k} (a_k a_j + a_k a_m) P_{ND} (k, j, m, i, l). 
\end{equation*} 
By repeating this rearrangement of terms also for $C_{01}$ and $C_{02}$, we can obtain \eqref{eq:K_ND} in the main text.

Now we are going to show that $\KND_{q,1} = 0$. To do this, we have to derive some properties of the ND quasi-probability distribution. 
Let us consider the case in which $m = j$ and $l \neq i$.
We want to show that $P_{ND}(k, k, k, i, \bar{i}) = - P_{ND}(\bar{k}, \bar{k}, \bar{k}, i, \bar{i})$. For a two-level system and binary observables, we have that $\hat{\Pi}_k = \Idoperator - \hat{\Pi}_{\bar{k}}$. Using this relation, by direct calculation, it holds that
\begin{eqnarray*}
     U_{2,kk} U^{*}_{2,kk} &=& 
     \Trace \left[ \hat{\Pi}_k \MU_2 \hat{\Pi}_k \MU_2^\dagger \right] = \Trace \left[ \hat{\Pi}_{\bar{k}} \MU_2 \hat{\Pi}_{\bar{k}} \MU_2^\dagger \right] \\
     &=& U_{2,\bar{k}\bar{k}} U^{*}_{2,\bar{k}\bar{k}}\,.
\end{eqnarray*}
Analogously, we get that 
\begin{eqnarray*}
     U_{1,ki} U^{*}_{1,  \bar{i} k} 
     &=& \bra{\bar{i}} \MU_1^\dagger ( \Idoperator - \hat{\Pi}_{\bar{k}}) \MU_1 \ket{i} = 
     - U^{*}_{1,\bar{i} \bar{k}} U_{1,\bar{k} i}.
\end{eqnarray*}
Hence, 
\begin{eqnarray*}
	P_{ND}(k, k, k, i, \bar{i}) &=&  U_{2,kk} U_{1,ki} \rho^0_{i \bar{i}} U^{*}_{1, \bar{i} k} U^{*}_{2,kk} \\ 
	&=&	 - U_{2,\bar{k}\bar{k}} U_{1,\bar{k} i} \rho^0_{i \bar{i}} U^{*}_{1,\bar{i} \bar{k}} U^{*}_{2, \bar{k}\bar{k}} \\
	&=& - P_{ND}(\bar{k}, \bar{k}, \bar{k}, i, \bar{i}). 
\end{eqnarray*}
In a similar way, it can be proven that $P_{ND}(k, \bar{k}, \bar{k}, i, \bar{i}) = -P_{ND}(\bar{k}, k, k, i, \bar{i})$.
At this point, let us take the expression of $\KND_{q,1}$ [\eqref{eq:K1_quant} in the main text]. Separating the contribution for $a_j=a_k$ and $a_j \neq a_k$ and recalling that $a_k^2=1$, $\KND_{q,1}$ can be equivalently written as
\begin{equation*}
 	\KND_{q,1} = \sum_{i} \left\{ \sum_{k} P_{ND}(k, k, k, i, \bar{i}) -  \sum_{k} P_{ND}(k, \bar{k}, \bar{k}, i, \bar{i}) \right \}.
\end{equation*}
Using the properties of the ND quasi-probability derived above, we have that both the sums vanish with the result that $\KND_{q,1} = 0$.

Let us now derive the simplified expression of $\KND_{q,2}$ as given by \eqref{eq:final_K_q2} in the main text. We are going to use the relation $\sum_{k} P_{ND}(k, j, \bar{j}, i, i) = 0$, whose proof is in the following $1$-line calculation:
\begin{eqnarray*}
	\sum_{k} P_{ND}(k, j, \bar{j}, i, i) &=& \sum_{k} \Trace \left[ \hat{\Pi}_k \MU_2 \hat{\Pi}_j \MU_1 \hat{\Pi}_i \hat{\rho}^0 \hat{\Pi}_i \MU_1^\dagger \hat{\Pi}_{\bar{j}} \MU_2^\dagger \right] \\
	&=& \sum_{k} \Trace \left [ \hat{\Pi}_j \MU_1 \hat{\Pi}_i \hat{\rho}^0 \hat{\Pi}_i \MU_1^\dagger \hat{\Pi}_{\bar{j}} \right] = 0.  
\end{eqnarray*}
From this equality, we get $P_{ND}(k, j, \bar{j}, i, i) = -P_{ND}(\bar{k}, j, \bar{j}, i, i)$.
Moreover, recalling that $P_{ND}(k, m, j, i, l) = ( P_{ND}(k, j, m, i, l) )^*$, we also have that $\sum_{j} P_{ND}(k, j, \bar{j}, i, i) = 2 \, \re \left[P_{ND}(k, j, \bar{j}, i, i) \right]$. As a result, the contribution $\KND_{q,2}$ in \eqref{eq:K2_quant} (with constraints $m=\bar{j}$ and $l=1$) simplifies to
\begin{eqnarray*}
 	\KND_{q,2} &=& - \sum_{i, j, k} a_k a_i   P_{ND}(k, j, \bar{j}, i, i) \\
	&=&
    - \sum_{j, k} \left\{ P_{ND}(k, j, \bar{j}, k, k) - P_{ND}(k, j, \bar{j}, \bar{k}, \bar{k}) \right\}
  \\
	&=& -2 \sum_{j} \left\{   P_{ND}(k, j, \bar{j}, k, k) + P_{ND}(\bar{k}, j, \bar{j}, \bar{k}, \bar{k})  \right\} \\
	&=& -4 \left\{  \re \left[ P_{ND}(k, j, \bar{j}, k, k) \right] + \re \left[ P_{ND}(\bar{k}, j, \bar{j}, \bar{k}, \bar{k}) \right]    \right\}  \\
    &=& -4 \sum_{k} \re \left[ P_{ND}(k, j, \bar{j}, k, k) \right]
\end{eqnarray*}
that corresponds to \eqref{eq:final_K_q2} in the main text. 

\subsection*{Acknowledgments}

The authors acknowledge the fruitful discussion with Andrea Smirne.
P.S.~acknowledges financial support from INFN. S.G.~would like to thank the PRIN project 2022FEXLYB Quantum Reservoir Computing (QuReCo), the PNRR MUR project PE0000023-NQSTI funded by the European Union--Next Generation EU, and the MISTI Global Seed Funds MIT-FVG Collaboration Grant ``Revealing and exploiting quantumness via quasiprobabilities: from quantum thermodynamics to quantum sensing''.

\subsection*{Authors' contributions}

P.S. designed and performed the research. P.S. and S.G. discussed the results and wrote the paper.

\bibliography{ND_vs_LG}

\begin{thebibliography}{49}
\providecommand{\natexlab}[1]{#1}
\providecommand{\url}[1]{\texttt{#1}}
\expandafter\ifx\csname urlstyle\endcsname\relax
  \providecommand{\doi}[1]{doi: #1}\else
  \providecommand{\doi}{doi: \begingroup \urlstyle{rm}\Url}\fi

\bibitem[Bell(1964)]{Bell1964}
J.~S. Bell.
\newblock {On the Einstein Podolsky Rosen paradox}.
\newblock \emph{Physics Physique Fizika}, 1:\penalty0 195--200, Nov 1964.
\newblock \doi{10.1103/PhysicsPhysiqueFizika.1.195}.
\newblock URL \url{https://link.aps.org/doi/10.1103/PhysicsPhysiqueFizika.1.195}.

\bibitem[Leggett and Garg(1985)]{Leggett1985}
A.~J. Leggett and Anupam Garg.
\newblock Quantum mechanics versus macroscopic realism: {I}s the flux there when nobody looks?
\newblock \emph{Phys. Rev. Lett.}, 54:\penalty0 857--860, Mar 1985.
\newblock \doi{10.1103/PhysRevLett.54.857}.

\bibitem[Freedman and Clauser(1972)]{Freedman1972}
Stuart~J. Freedman and John~F. Clauser.
\newblock {Experimental Test of Local Hidden-Variable Theories}.
\newblock \emph{Phys. Rev. Lett.}, 28:\penalty0 938--941, Apr 1972.
\newblock \doi{10.1103/PhysRevLett.28.938}.
\newblock URL \url{https://link.aps.org/doi/10.1103/PhysRevLett.28.938}.

\bibitem[Aspect et~al.(1982)Aspect, Grangier, and Roger]{Aspect1982}
Alain Aspect, Philippe Grangier, and G\'erard Roger.
\newblock {Experimental Realization of Einstein-Podolsky-Rosen-Bohm Gedankenexperiment: A New Violation of Bell's Inequalities}.
\newblock \emph{Phys. Rev. Lett.}, 49:\penalty0 91--94, Jul 1982.
\newblock \doi{10.1103/PhysRevLett.49.91}.
\newblock URL \url{https://link.aps.org/doi/10.1103/PhysRevLett.49.91}.

\bibitem[Pan et~al.(1998)Pan, Bouwmeester, Weinfurter, and Zeilinger]{Pan1998}
Jian-Wei Pan, Dik Bouwmeester, Harald Weinfurter, and Anton Zeilinger.
\newblock {Experimental Entanglement Swapping: Entangling Photons That Never Interacted}.
\newblock \emph{Phys. Rev. Lett.}, 80:\penalty0 3891--3894, May 1998.
\newblock \doi{10.1103/PhysRevLett.80.3891}.
\newblock URL \url{https://link.aps.org/doi/10.1103/PhysRevLett.80.3891}.

\bibitem[Maroney and Timpson(2014)]{maroney2014}
Owen~JE Maroney and Christopher~G Timpson.
\newblock {Quantum-vs. macro-realism: What does the Leggett-Garg inequality actually test?}
\newblock \emph{arXiv preprint arXiv:1412.6139}, 2014.

\bibitem[Schmid(2024)]{Schmid2024}
David Schmid.
\newblock A review and reformulation of macroscopic realism: resolving its deficiencies using the framework of generalized probabilistic theories.
\newblock \emph{{Quantum}}, 8:\penalty0 1217, January 2024.
\newblock ISSN 2521-327X.
\newblock \doi{10.22331/q-2024-01-03-1217}.
\newblock URL \url{https://doi.org/10.22331/q-2024-01-03-1217}.

\bibitem[Halliwell(2016)]{Halliwell2016}
J.~J. Halliwell.
\newblock {Leggett-Garg inequalities and no-signaling in time: A quasiprobability approach}.
\newblock \emph{Phys. Rev. A}, 93:\penalty0 022123, Feb 2016.
\newblock \doi{10.1103/PhysRevA.93.022123}.
\newblock URL \url{https://link.aps.org/doi/10.1103/PhysRevA.93.022123}.

\bibitem[Braginsky et~al.(1980)Braginsky, Vorontsov, and Thorne]{Braginsky1980}
Vladimir~B. Braginsky, Yuri~I. Vorontsov, and Kip~S. Thorne.
\newblock {Quantum Nondemolition Measurements}.
\newblock \emph{Science}, 209\penalty0 (4456):\penalty0 547--557, 1980.
\newblock \doi{10.1126/science.209.4456.547}.
\newblock URL \url{https://www.science.org/doi/abs/10.1126/science.209.4456.547}.

\bibitem[Braginski{\u \i} et~al.(1992)Braginski{\u \i}, Khalili, and Thorne]{BraginskyBook}
V.~B. Braginski{\u \i}, Farid~Ya. Khalili, and Kip~S. Thorne.
\newblock \emph{Quantum measurement}.
\newblock Cambridge University Press, Cambridge [England] ;, 1992.
\newblock ISBN 052141928X.

\bibitem[Caves(1980)]{Caves1980}
Carlton~M. Caves.
\newblock {Quantum-Mechanical Radiation-Pressure Fluctuations in an Interferometer}.
\newblock \emph{Phys. Rev. Lett.}, 45:\penalty0 75--79, Jul 1980.
\newblock \doi{10.1103/PhysRevLett.45.75}.
\newblock URL \url{https://link.aps.org/doi/10.1103/PhysRevLett.45.75}.

\bibitem[Caves et~al.(1980)Caves, Thorne, Drever, Sandberg, and Zimmermann]{CavesRevModPhys}
Carlton~M. Caves, Kip~S. Thorne, Ronald W.~P. Drever, Vernon~D. Sandberg, and Mark Zimmermann.
\newblock {On the measurement of a weak classical force coupled to a quantum-mechanical oscillator. I. Issues of principle}.
\newblock \emph{Rev. Mod. Phys.}, 52:\penalty0 341--392, Apr 1980.
\newblock \doi{10.1103/RevModPhys.52.341}.
\newblock URL \url{https://link.aps.org/doi/10.1103/RevModPhys.52.341}.

\bibitem[Solinas et~al.(2013)Solinas, Averin, and Pekola]{solinas2013work}
Paolo Solinas, Dmitri~V. Averin, and Jukka~P. Pekola.
\newblock Work and its fluctuations in a driven quantum system.
\newblock \emph{Phys. Rev. B}, 87:\penalty0 060508, Feb 2013.
\newblock \doi{10.1103/PhysRevB.87.060508}.
\newblock URL \url{http://link.aps.org/doi/10.1103/PhysRevB.87.060508}.

\bibitem[Solinas and Gasparinetti(2015)]{solinas2015fulldistribution}
P.~Solinas and S.~Gasparinetti.
\newblock Full distribution of work done on a quantum system for arbitrary initial states.
\newblock \emph{Phys. Rev. E}, 92:\penalty0 042150, Oct 2015.
\newblock \doi{10.1103/PhysRevE.92.042150}.
\newblock URL \url{http://link.aps.org/doi/10.1103/PhysRevE.92.042150}.

\bibitem[Solinas and Gasparinetti(2016)]{solinas2016probing}
P.~Solinas and S.~Gasparinetti.
\newblock Probing quantum interference effects in the work distribution.
\newblock \emph{Phys. Rev. A}, 94:\penalty0 052103, Nov 2016.
\newblock \doi{10.1103/PhysRevA.94.052103}.
\newblock URL \url{https://link.aps.org/doi/10.1103/PhysRevA.94.052103}.

\bibitem[Solinas et~al.(2021)Solinas, Amico, and Zangh\`{\i}]{solinas2021}
P.~Solinas, M.~Amico, and N.~Zangh\`{\i}.
\newblock Measurement of work and heat in the classical and quantum regimes.
\newblock \emph{Phys. Rev. A}, 103:\penalty0 L060202, Jun 2021.
\newblock \doi{10.1103/PhysRevA.103.L060202}.
\newblock URL \url{https://link.aps.org/doi/10.1103/PhysRevA.103.L060202}.

\bibitem[Solinas et~al.(2022)Solinas, Amico, and Zangh\`{\i}]{solinas2022}
P.~Solinas, M.~Amico, and N.~Zangh\`{\i}.
\newblock Quasiprobabilities of work and heat in an open quantum system.
\newblock \emph{Phys. Rev. A}, 105:\penalty0 032606, Mar 2022.
\newblock \doi{10.1103/PhysRevA.105.032606}.
\newblock URL \url{https://link.aps.org/doi/10.1103/PhysRevA.105.032606}.

\bibitem[Gherardini and De~Chiara(2024)]{GherardiniTutorial}
Stefano Gherardini and Gabriele De~Chiara.
\newblock {Quasiprobabilities in quantum thermodynamics and many-body systems: A tutorial}.
\newblock \emph{arXiv preprint arXiv:2403.17138}, 2024.
\newblock \doi{10.48550/arXiv.2403.17138}.
\newblock URL \url{https://doi.org/10.48550/arXiv.2403.17138}.

\bibitem[Lostaglio et~al.(2023)Lostaglio, Belenchia, Levy, Hern\'andez-G\'omez, Fabbri, and Gherardini]{LostaglioKirkwood2022}
Matteo Lostaglio, Alessio Belenchia, Amikam Levy, Santiago Hern\'andez-G\'omez, Nicole Fabbri, and Stefano Gherardini.
\newblock {Kirkwood-Dirac quasiprobability approach to the statistics of incompatible observables}.
\newblock \emph{Quantum}, 7:\penalty0 1128, 2023.
\newblock \doi{10.22331/q-2023-10-09-1128}.
\newblock URL \url{https://quantum-journal.org/papers/q-2023-10-09-1128/}.

\bibitem[Yunger~Halpern et~al.(2018)Yunger~Halpern, Swingle, and Dressel]{yunger2018quasiprobability}
N.~Yunger~Halpern, B.~Swingle, and J.~Dressel.
\newblock Quasiprobability behind the out-of-time-ordered correlator.
\newblock \emph{Phys. Rev. A}, 97:\penalty0 042105, Apr 2018.
\newblock \doi{10.1103/PhysRevA.97.042105}.
\newblock URL \url{https://link.aps.org/doi/10.1103/PhysRevA.97.042105}.

\bibitem[Arvidsson-Shukur et~al.(2021)Arvidsson-Shukur, Chevalier~Drori, and Yunger~Halpern]{ArvidssonShukurJPA2021}
D.~R.~M. Arvidsson-Shukur, J.~Chevalier~Drori, and N.~Yunger~Halpern.
\newblock Conditions tighter than noncommutation needed for nonclassicality.
\newblock \emph{J. Phys. A: Math. Theor.}, 54:\penalty0 284001, 2021.
\newblock \doi{10.1088/1751-8121/ac0289}.
\newblock URL \url{https://iopscience.iop.org/article/10.1088/1751-8121/ac0289}.

\bibitem[De~Bi\`evre(2021)]{DeBievrePRL2021}
S.~De~Bi\`evre.
\newblock {Complete Incompatibility, Support Uncertainty, and Kirkwood-Dirac Nonclassicality}.
\newblock \emph{Phys. Rev. Lett.}, 127:\penalty0 190404, Nov 2021.
\newblock \doi{10.1103/PhysRevLett.127.190404}.
\newblock URL \url{https://link.aps.org/doi/10.1103/PhysRevLett.127.190404}.

\bibitem[Budiyono and Dipojono(2023)]{BudiyonoPRAquantifying}
A.~Budiyono and H.~K. Dipojono.
\newblock {Quantifying quantum coherence via Kirkwood-Dirac quasiprobability}.
\newblock \emph{Phys. Rev. A}, 107:\penalty0 022408, Feb 2023.
\newblock \doi{10.1103/PhysRevA.107.022408}.
\newblock URL \url{https://link.aps.org/doi/10.1103/PhysRevA.107.022408}.

\bibitem[Wagner et~al.(2024)Wagner, Schwartzman-Nowik, Paiva, Te'eni, Ruiz-Molero, Barbosa, Cohen, and Galv{\~a}o]{wagner2023quantum}
R.~Wagner, Z.~Schwartzman-Nowik, I.~L. Paiva, A.~Te'eni, A.~Ruiz-Molero, R.~Soares Barbosa, E.~Cohen, and E.~F. Galv{\~a}o.
\newblock {Quantum circuits for measuring weak values, Kirkwood--Dirac quasiprobability distributions, and state spectra}.
\newblock \emph{Quantum Sci. Technol.}, 9:\penalty0 015030, 2024.
\newblock \doi{10.1088/2058-9565/ad124c}.
\newblock URL \url{https://iopscience.iop.org/article/10.1088/2058-9565/ad124c}.

\bibitem[Arvidsson-Shukur et~al.(2024)Arvidsson-Shukur, Braasch~Jr., De~Bievre, Dressel, Jordan, Langrenez, Lostaglio, Lundeen, and Yunger~Halpern]{ArvidssonShukur2024review}
David R.~M. Arvidsson-Shukur, William~F. Braasch~Jr., Stephan De~Bievre, Justin Dressel, Andrew~N. Jordan, Christopher Langrenez, Matteo Lostaglio, Jeff~S. Lundeen, and Nicole Yunger~Halpern.
\newblock {Properties and Applications of the Kirkwood-Dirac Distribution}.
\newblock \emph{arXiv preprint arXiv:2403.18899}, 2024.
\newblock \doi{10.48550/arXiv.2403.18899}.
\newblock URL \url{https://doi.org/10.48550/arXiv.2403.18899}.

\bibitem[Hern{\'a}ndez-G{\'o}mez et~al.(2024)Hern{\'a}ndez-G{\'o}mez, Isogawa, Belenchia, Levy, Fabbri, Gherardini, and Cappellaro]{hernandezArXiv2024Interfero}
Santiago Hern{\'a}ndez-G{\'o}mez, Takuya Isogawa, Alessio Belenchia, Amikam Levy, Nicole Fabbri, Stefano Gherardini, and Paola Cappellaro.
\newblock Interferometry of quantum correlation functions to access quasiprobability distribution of work.
\newblock \emph{arXiv preprint arXiv:2405.21041}, 2024.
\newblock URL \url{https://arxiv.org/abs/2405.21041}.

\bibitem[Levitov et~al.(1996)Levitov, Lee, and Lesovik]{LevitovJMP1996}
Leonid~S. Levitov, Hyunwoo Lee, and Gordey~B. Lesovik.
\newblock Electron counting statistics and coherent states of electric current.
\newblock \emph{J. Math. Phys.}, 37\penalty0 (10):\penalty0 4845--4866, 1996.
\newblock \doi{10.1063/1.531672}.
\newblock URL \url{https://pubs.aip.org/aip/jmp/article-abstract/37/10/4845/454669/Electron-counting-statistics-and-coherent-states?redirectedFrom=fulltext}.

\bibitem[Nazarov and Kindermann(2003)]{NazarovEPJB2003}
Yu~V. Nazarov and M.~Kindermann.
\newblock Full counting statistics of a general quantum mechanical variable.
\newblock \emph{Eur. Phys. J. B}, 35:\penalty0 413--420, 2003.
\newblock \doi{10.1140/epjb/e2003-00293-1}.
\newblock URL \url{https://link.springer.com/article/10.1140/epjb/e2003-00293-1}.

\bibitem[Clerk(2011)]{ClerkPRA2011}
A.~A. Clerk.
\newblock Full counting statistics of energy fluctuations in a driven quantum resonator.
\newblock \emph{Phys. Rev. A}, 84:\penalty0 043824, Oct 2011.
\newblock \doi{10.1103/PhysRevA.84.043824}.
\newblock URL \url{https://link.aps.org/doi/10.1103/PhysRevA.84.043824}.

\bibitem[Hofer and Clerk(2016)]{HoferPRL2016}
Patrick~P. Hofer and A.~A. Clerk.
\newblock {Negative Full Counting Statistics Arise from Interference Effects}.
\newblock \emph{Phys. Rev. Lett.}, 116:\penalty0 013603, Jan 2016.
\newblock \doi{10.1103/PhysRevLett.116.013603}.
\newblock URL \url{https://link.aps.org/doi/10.1103/PhysRevLett.116.013603}.

\bibitem[Hofer(2017)]{Hofer2017quasiprobability}
Patrick~P. Hofer.
\newblock Quasi-probability distributions for observables in dynamic systems.
\newblock \emph{{Quantum}}, 1:\penalty0 32, October 2017.
\newblock ISSN 2521-327X.
\newblock \doi{10.22331/q-2017-10-12-32}.
\newblock URL \url{https://doi.org/10.22331/q-2017-10-12-32}.

\bibitem[Potts(2019)]{PottsPRL2019}
Patrick~P. Potts.
\newblock {Certifying Nonclassical Behavior for Negative Keldysh Quasiprobabilities}.
\newblock \emph{Phys. Rev. Lett.}, 122:\penalty0 110401, Mar 2019.
\newblock \doi{10.1103/PhysRevLett.122.110401}.
\newblock URL \url{https://link.aps.org/doi/10.1103/PhysRevLett.122.110401}.

\bibitem[Ruskov et~al.(2006)Ruskov, Korotkov, and Mizel]{Ruskov2006}
Rusko Ruskov, Alexander~N. Korotkov, and Ari Mizel.
\newblock Signatures of quantum behavior in single-qubit weak measurements.
\newblock \emph{Phys. Rev. Lett.}, 96:\penalty0 200404, May 2006.
\newblock \doi{10.1103/PhysRevLett.96.200404}.
\newblock URL \url{https://link.aps.org/doi/10.1103/PhysRevLett.96.200404}.

\bibitem[Jordan et~al.(2006)Jordan, Korotkov, and B\"uttiker]{Jordan2006}
Andrew~N. Jordan, Alexander~N. Korotkov, and Markus B\"uttiker.
\newblock Leggett-garg inequality with a kicked quantum pump.
\newblock \emph{Phys. Rev. Lett.}, 97:\penalty0 026805, Jul 2006.
\newblock \doi{10.1103/PhysRevLett.97.026805}.
\newblock URL \url{https://link.aps.org/doi/10.1103/PhysRevLett.97.026805}.

\bibitem[Goggin et~al.(2011)Goggin, Almeida, Barbieri, Lanyon, O'Brien, White, and Pryde]{Goggin2011}
M.~E. Goggin, M.~P. Almeida, M.~Barbieri, B.~P. Lanyon, J.~L. O'Brien, A.~G. White, and G.~J. Pryde.
\newblock Violation of the leggett--garg inequality with weak measurements of photons.
\newblock \emph{Proceedings of the National Academy of Sciences}, 108\penalty0 (4):\penalty0 1256--1261, 2011.
\newblock \doi{10.1073/pnas.1005774108}.
\newblock URL \url{https://www.pnas.org/doi/abs/10.1073/pnas.1005774108}.

\bibitem[Arndt et~al.(1999)Arndt, Nairz, Vos-Andreae, Keller, van~der Zouw, and Zeilinger]{Arndt1999}
Markus Arndt, Olaf Nairz, Julian Vos-Andreae, Claudia Keller, Gerbrand van~der Zouw, and Anton Zeilinger.
\newblock {Wave--particle duality of C60 molecules}.
\newblock \emph{Nature}, 401\penalty0 (6754):\penalty0 680--682, 1999.
\newblock \doi{10.1038/44348}.
\newblock URL \url{https://doi.org/10.1038/44348}.

\bibitem[Gerlich et~al.(2011)Gerlich, Eibenberger, Tomandl, Nimmrichter, Hornberger, Fagan, T{\"u}xen, Mayor, and Arndt]{Gerlich2011}
Stefan Gerlich, Sandra Eibenberger, Mathias Tomandl, Stefan Nimmrichter, Klaus Hornberger, Paul~J. Fagan, Jens T{\"u}xen, Marcel Mayor, and Markus Arndt.
\newblock Quantum interference of large organic molecules.
\newblock \emph{Nature Communications}, 2\penalty0 (1):\penalty0 263, 2011.
\newblock \doi{10.1038/ncomms1263}.
\newblock URL \url{https://doi.org/10.1038/ncomms1263}.

\bibitem[Aspelmeyer(2022)]{Aspelmeyer2022}
Markus Aspelmeyer.
\newblock \emph{{When Zeh Meets Feynman: How to Avoid the Appearance of a Classical World in Gravity Experiments}}, pages 85--95.
\newblock Springer International Publishing, Cham, 2022.
\newblock ISBN 978-3-030-88781-0.
\newblock \doi{10.1007/978-3-030-88781-0_5}.
\newblock URL \url{https://doi.org/10.1007/978-3-030-88781-0_5}.

\bibitem[Fuchs et~al.(2024)Fuchs, Uitenbroek, Plugge, van Halteren, van Soest, Vinante, Ulbricht, and Oosterkamp]{FuchsScienceAdvances2024}
Tim~M. Fuchs, Dennis~G. Uitenbroek, Jaimy Plugge, Noud van Halteren, Jean-Paul van Soest, Andrea Vinante, Hendrik Ulbricht, and Tjerk~H. Oosterkamp.
\newblock {Measuring gravity with milligram levitated masses}.
\newblock \emph{Science Advances}, 10:\penalty0 1--6, 2024.
\newblock \doi{10.1126/sciadv.adk2949}.

\bibitem[Kim et~al.(2023)Kim, Eddins, Anand, Wei, van~den Berg, Rosenblatt, Nayfeh, Wu, Zaletel, Temme, and Kandala]{Kim2023}
Youngseok Kim, Andrew Eddins, Sajant Anand, Ken~Xuan Wei, Ewout van~den Berg, Sami Rosenblatt, Hasan Nayfeh, Yantao Wu, Michael Zaletel, Kristan Temme, and Abhinav Kandala.
\newblock Evidence for the utility of quantum computing before fault tolerance.
\newblock \emph{Nature}, 618\penalty0 (7965):\penalty0 500--505, 2023.
\newblock \doi{10.1038/s41586-023-06096-3}.
\newblock URL \url{https://doi.org/10.1038/s41586-023-06096-3}.

\bibitem[Bravyi et~al.(2024)Bravyi, Cross, Gambetta, Maslov, Rall, and Yoder]{Bravyi2024}
Sergey Bravyi, Andrew~W. Cross, Jay~M. Gambetta, Dmitri Maslov, Patrick Rall, and Theodore~J. Yoder.
\newblock High-threshold and low-overhead fault-tolerant quantum memory.
\newblock \emph{Nature}, 627\penalty0 (8005):\penalty0 778--782, 2024.
\newblock \doi{10.1038/s41586-024-07107-7}.
\newblock URL \url{https://doi.org/10.1038/s41586-024-07107-7}.

\bibitem[Nath et~al.(2024)Nath, Saha, Home, and Sinha]{Nath2024}
Pingal~Pratyush Nath, Debashis Saha, Dipankar Home, and Urbasi Sinha.
\newblock {Single-System-Based Generation of Certified Randomness Using Leggett-Garg Inequality}.
\newblock \emph{Phys. Rev. Lett.}, 133:\penalty0 020802, Jul 2024.
\newblock \doi{10.1103/PhysRevLett.133.020802}.
\newblock URL \url{https://link.aps.org/doi/10.1103/PhysRevLett.133.020802}.

\bibitem[Knee et~al.(2016)Knee, Kakuyanagi, Yeh, Matsuzaki, Toida, Yamaguchi, Saito, Leggett, and Munro]{Knee2016}
George~C. Knee, Kosuke Kakuyanagi, Mao-Chuang Yeh, Yuichiro Matsuzaki, Hiraku Toida, Hiroshi Yamaguchi, Shiro Saito, Anthony~J. Leggett, and William~J. Munro.
\newblock A strict experimental test of macroscopic realism in a superconducting flux qubit.
\newblock \emph{Nature Communications}, 7\penalty0 (1):\penalty0 13253, 2016.
\newblock \doi{10.1038/ncomms13253}.
\newblock URL \url{https://doi.org/10.1038/ncomms13253}.

\bibitem[Kofler and Brukner(2008)]{Kofler2008}
Johannes Kofler and \ifmmode \check{C}\else~\v{C}\fi{}aslav Brukner.
\newblock {Conditions for Quantum Violation of Macroscopic Realism}.
\newblock \emph{Phys. Rev. Lett.}, 101:\penalty0 090403, Aug 2008.
\newblock \doi{10.1103/PhysRevLett.101.090403}.
\newblock URL \url{https://link.aps.org/doi/10.1103/PhysRevLett.101.090403}.

\bibitem[Solinas et~al.(2017)Solinas, Miller, and Anders]{solinasPRA2017}
P.~Solinas, H.~J.~D. Miller, and J.~Anders.
\newblock Measurement-dependent corrections to work distributions arising from quantum coherences.
\newblock \emph{Phys. Rev. A}, 96:\penalty0 052115, Nov 2017.
\newblock \doi{10.1103/PhysRevA.96.052115}.
\newblock URL \url{https://link.aps.org/doi/10.1103/PhysRevA.96.052115}.

\bibitem[De~Chiara et~al.(2018)De~Chiara, Solinas, Cerisola, and Roncaglia]{DeChiara2018}
Gabriele De~Chiara, Paolo Solinas, Federico Cerisola, and Augusto~J. Roncaglia.
\newblock \emph{Ancilla-Assisted Measurement of Quantum Work}, pages 337--362.
\newblock Springer International Publishing, Cham, 2018.
\newblock ISBN 978-3-319-99046-0.
\newblock \doi{10.1007/978-3-319-99046-0_14}.
\newblock URL \url{https://doi.org/10.1007/978-3-319-99046-0_14}.

\bibitem[Fritz(2010)]{Fritz_2010}
Tobias Fritz.
\newblock {Quantum correlations in the temporal Clauser--Horne--Shimony--Holt (CHSH) scenario}.
\newblock \emph{New Journal of Physics}, 12\penalty0 (8):\penalty0 083055, aug 2010.
\newblock \doi{10.1088/1367-2630/12/8/083055}.
\newblock URL \url{https://dx.doi.org/10.1088/1367-2630/12/8/083055}.

\bibitem[Kofler and Brukner(2013)]{Kofler2013}
Johannes Kofler and \ifmmode \check{C}\else~\v{C}\fi{}aslav Brukner.
\newblock {Condition for macroscopic realism beyond the Leggett-Garg inequalities}.
\newblock \emph{Phys. Rev. A}, 87:\penalty0 052115, May 2013.
\newblock \doi{10.1103/PhysRevA.87.052115}.
\newblock URL \url{https://link.aps.org/doi/10.1103/PhysRevA.87.052115}.

\bibitem[Emary et~al.(2013)Emary, Lambert, and Nori]{Emary_2014}
Clive Emary, Neill Lambert, and Franco Nori.
\newblock {Leggett--Garg inequalities}.
\newblock \emph{Reports on Progress in Physics}, 77\penalty0 (1):\penalty0 016001, dec 2013.
\newblock \doi{10.1088/0034-4885/77/1/016001}.
\newblock URL \url{https://dx.doi.org/10.1088/0034-4885/77/1/016001}.

\end{thebibliography}

\end{document}